\documentclass[aps,preprint,onecolumn,floatfix,superscriptaddress,nofootinbib]{revtex4-1}

\usepackage{amsmath, amssymb}
\usepackage{graphicx,bm}
\usepackage{multirow}
\usepackage{makecell}
\usepackage{slashed}
\usepackage{cancel}
\usepackage{booktabs} 
\usepackage{float}    
\usepackage[colorlinks,citecolor=blue]{hyperref}
\usepackage[caption=false]{subfig}

\begin{document}
\title{Current status and prospects of light bino-higgsino dark matter in natural SUSY}

\author{XinTian Wang}
\affiliation{School of Physical Science and Technology, Xinjiang University,Urumqi 830046, China}
\altaffiliation{}

\author{Murat Abdughani}
\email{mulati@xju.edu.cn}
\affiliation{School of Physical Science and Technology, Xinjiang University,Urumqi 830046, China}

\begin{abstract}
Given recent advancements in dark matter (DM) search experiments, particularly the latest LUX-ZEPLIN (LZ) direct detection (DD) results, we systematically investigate the light bino--higgsino DM scenario within the natural supersymmetric framework. Requiring the electroweak fine-tuning parameter $\Delta_{\mathrm{EW}} < 30$ fixes the higgsino mass parameter in the range of $|\mu| \in [100, 350]$~GeV, while we extend the bino mass to $M_1 \in [10, 350]$~GeV. Incorporating constraints from Higgs physics, rare $B$ decays, LEP limits, and DD experiments, we find that part of the parameter space remains viable. However, the relic density of neutralino DM necessarily lies below the observed Planck value, contributing at most $\sim$2\% of the total DM abundance. Some of the surviving parameter space is already excluded by current 1\mbox{3 TeV LHC sear}ches, while the future 14 TeV HL-LHC with 3000 fb$^{-1}$ luminosity will probe the remaining region of the considered parameter space.
\end{abstract}
 
\date{\today}
\maketitle
\newpage

\section{Introduction}

Cosmological observations provide compelling evidence for the existence of dark matter (DM), yet its particle nature remains one of the foremost challenges in particle physics and cosmology~\cite{Planck:2018vyg}. Additionally, the discovery of the Higgs boson at the Large Hadron Collider (LHC)~\cite{ATLAS:2012yve,CMS:2012qbp} has intensified the need to address the hierarchy problem, which arises from the large disparity between the electroweak scale and the Planck scale, necessitating fine-tuned cancellations to maintain the observed Higgs mass. This issue challenges the naturalness of the Standard Model (SM), motivating the exploration of new physics models that stabilize the electroweak scale. Weak-scale supersymmetry (SUSY), particularly the Minimal Supersymmetric SM (MSSM), is a leading framework to address these issues at the TeV scale~\cite{Martin:1997ns}. The MSSM resolves the hierarchy problem by introducing SUSY, which cancels quadratic divergences in the Higgs mass, thereby enhancing naturalness~\cite{JUNGMAN1996195}. It also provides a compelling dark matter candidate in the form of the lightest supersymmetric particle (LSP)~\cite{PhysRevD.105.115029,Baer_2020}, stabilized by R-parity~\cite{Kohler:2024mpu,Nangia:2024ghc} conservation.

Within this framework, natural SUSY is a well-motivated approach that minimizes fine-tuning by predicting light stop and higgsinos with masses near the electroweak scale while allowing heavier masses for other superpartners~\cite{Papucci:2011wy,Baer:2012up,Boehm:1994gt,Kitano:2006gv,Hall:2011jd}. This hierarchy aligns with the observed Higgs mass of approximately 125~GeV and remains consistent with LHC constraints~\cite{ATLAS:2012yve,CMS:2012qbp}. In natural SUSY, the LSP is typically a neutralino, and a light higgsino-like LSP is common due to the small $\mu$ parameter required for low fine-tuning. However, a pure higgsino LSP often yields a thermal relic density below the observed value of $\Omega h^2 \approx 0.12$ due to its large annihilation rate~\cite{ArkaniHamed:2006mb,Martin:2024ytt}. Mixed bino--higgsino neutralino as the LSP in natural SUSY allow a light bino by relaxing the gaugino mass unification assumption. This scenario achieves the correct relic density through enhanced annihilation channels mediated by higgsino components~\cite{ArkaniHamed:2006mb,CahillRowley:2012cb,Han:2013usa,Huang:2014xua,Baer:2016ucr,Abdughani:2017dqs,Abdughani:2019wss,Boehm:2013qva,Cao:2015efs,Guchait:2021tmh,Chatterjee:2025gej}. The light bino--higgsino scenario has been extensively studied in the literature~\cite{Abdughani:2017dqs, Abdughani:2019wss, Boehm:2013qva, Cao:2015efs, Guchait:2021tmh, Chatterjee:2025gej}. Previous studies have thoroughly mapped out its relic density mechanisms—highlighting the roles of coannihilation and resonant funnels—and demonstrated how theoretical blind spots could suppress direct detection cross secti\mbox{ons~\cite{coann,coann1,resonance1,resonance2,coannihilate}}. Current direct detection (DD) experiments, especially the limits reported by the 2025 LUX-ZEPLIN (LZ) results~\cite{LZ:2024zvo}, impose stringent constraints on the weakly interacting massive particle (WIMP) parameter space. LHC searches for electroweakinos also strongly constrain this parameter space. Indirect detection experiments are not effective in this region since the annihilation is $p$-wave-suppressed.

In this work, we investigate the light bino--higgsino dark matter scenario in natural SUSY, focusing on its compatibility with recent DD constraints from LZ and LHC. We explore the light dark matter regime ($m_{\tilde{\chi}^0_1} < 350$~GeV) and determine the parameter space limit for the LSP that constitutes all or some portion of DM, i.e., DM relic density can be smaller than the observed value. The small $\mu$ parameter in natural SUSY enhances bino--higgsino mixing, impacting the spin-independent (SI) and spin-dependent (SD) neutralino--nucleon scattering cross sections. Additionally, we evaluate the potential to probe this scenario through electroweakino searches at the 14~TeV LHC.

The paper is organized as follows: in Section~\ref{sec:neutralino}, we discuss the bino--higgsino neutralino parameter space in natural SUSY; in Section~\ref{sec:likelihood}, we present our parameter scan regions, adopted constraints, and likelihoods; in Section~\ref{sec:result}, we give obtained results; and in Section~\ref{sec:conclusion}, we summarize our conclusions.

\section{Light neutralino DM} \label{sec:neutralino}

The two neutral higgsinos ($\tilde{H}_u^0$ and $\tilde{H}_d^0$) and the two neutral gauginos ($\tilde{B}$ and $\tilde{W}^0$) are combined to form four mass eigenstates called neutralinos. In the gauge-eigenstate basis ($\tilde{B}$, $\tilde{W}^0$, $\tilde{H}_d$, $\tilde{H}_u$), the neutralino mass matrix takes the form: 
\begin{equation}
M_{\tilde{\chi}^0} = \begin{pmatrix}
M_1 & 0 & -c_\beta s_W m_Z & s_\beta s_W m_Z \\
0 & M_2 & c_\beta c_W m_Z  & -s_\beta s_W m_Z \\
-c_\beta s_W m_Z & c_\beta c_W m_Z & 0 & -\mu \\
s_\beta s_W m_Z & -s_\beta s_W m_Z & -\mu & 0
\end{pmatrix},
\label{eq:neutralinomatrix}
\end{equation}
where $s_\beta = \sin\beta$, $c_\beta = \cos\beta$, $s_W = \sin\theta_W$, $c_W = \cos\theta_W$, $M_1$ and $M_2$ are the soft-breaking mass parameters for bino and wino, respectively. $M_{\tilde{\chi}^0}$ can be diagonalized by a \mbox{$4\times 4$ unitary} matrix. In the limit of $M_1<\mu \ll M_2$, the lightest neutralino is bino-like (with some higgsino mixture), while the second and third neutralinos are higgsino-like. The LSP interacts with nuclei via the exchange of squarks and Higgs bosons (SI scattering) and via $Z$ boson and squarks exchange (SD scattering).

We introduced the problems that SUSY models can address, one of which is the widely discussed naturalness problem. In SUSY models, the mass of the $Z$ boson can be explained~\cite{PhysRevD.46.3981} by
\begin{equation}
\frac{M_Z^2}{2} = \frac{(m_{H_d}^2 + \Sigma_d) - (m_{H_u}^2 + \Sigma_u) \tan^2 \beta}{\tan^2 \beta - 1} - \mu^2,
\label{equation2}
\end{equation}
where $m_{H_d}^2$ and $m_{H_u}^2$ represent the soft symmetry-breaking parameters for the electroweak scale fields $H_d$ and $H_u$, respectively. $\Sigma_d$ and $\Sigma_u$ correspond to their radiative corrections. $\mu$ is the mass parameter associated with the higgsino, and $\tan \beta = v_u/v_d$. Equation~\eqref{equation2} is derived from the Higgs potential in the weak-scale MSSM, with all parameters evaluated at the scale $Q = M_{\mathrm{SUSY}}$. The observed values are obtained under the condition that there is no large cancellation between the terms on the right-hand side of Equation~\eqref{equation2}, meaning that none of these terms is larger than $M_Z^2$ in magnitude. Electroweak fine-tuning can be quantified by~\cite{Baer:2012cf}
\begin{equation}
     \Delta_{\mathrm{EW}} \equiv \max_i |C_i| / \left( \frac{M_Z^2}{2} \right),
\end{equation}
with $C_{H_d}=m_{H_d}^2/(\tan^2\beta -1)$, $C_{H_u}=-m_{H_u}^2\tan^2\beta /(\tan^2\beta -1)$, and $C_\mu =-\mu^2$. Also, $C_{\Sigma_u^u(k)} =-\Sigma_u^u(k)\tan^2\beta /(\tan^2\beta -1)$ and $C_{\Sigma_d^d(k)}=\Sigma_d^d(k)/(\tan^2\beta -1)$, where $k$ labels the various loop contributions included in Equation~\eqref{equation2}. 

A fully comprehensive evaluation of natural SUSY as a global framework requires minimizing both the tree-level contributions and these loop-level radiative corrections (particularly from stops and gluinos). However, the higgsino mass parameter $\mu$ is unique, as it governs the tree-level fine-tuning condition, representing the absolute irreducible floor of naturalness; any upper bound on $\Delta_{\mathrm{EW}}$ from electroweak naturalness considerations sets a strict corresponding limit on $\mu^2$~\cite{Chan:1997bi}. Imposing the upper limit of $\Delta_{\mathrm{EW}} < 30$~\cite{Das_2018,Ross_2016,Baer_2013} leads to the limit of $|\mu| \lesssim 350$~GeV~\cite{Tata:2020afe}. In this study, our primary focus is the dark matter phenomenology of the electroweakino sector. Therefore, while we strictly enforce this tree-level naturalness condition on the higgsinos, we adopt a phenomenological decoupling limit for the rest of the spectrum. By fixing the squark, slepton, and gluino mass parameters to 3~TeV, we ensure robust consistency with the observed 125~GeV Higgs mass and evade the severe constraints from LHC direct searches for colored superpartners. Furthermore, within the framework of radiatively driven natural supersymmetry, it has been demonstrated that multi-TeV stops and gluinos can still yield low electroweak fine-tuning ($\Delta_{\mathrm{EW}} < 30$) by minimizing the loop-level radiative corrections $\Sigma_u^u$~\cite{Baer_2013,Tata:2020afe}. Thus, this approach cleanly isolates the light bino--higgsino dark matter dynamics while remaining firmly within the established natural SUSY paradigm.

\section{Prior, constraints and likelihoods}
\label{sec:likelihood}

In our numerical calculations, we vary the relevant parameters in the ranges of
\begin{eqnarray}
&&100~\mathrm{GeV} \le |\mu| \le 350~\mathrm{GeV},\quad  10~\mathrm{GeV}  \le M_1 \le 350~\mathrm{GeV},\nonumber \\ 
&&|A_T| \le 4000~\mathrm{GeV},\quad  5 \le \tan\beta \le 50,\quad \mathrm{others} =  3~\mathrm{TeV}.
\label{eq:range}
\end{eqnarray}

The lower limit on $\mu$ comes from LEP charged particle searches~\cite{ALEPH:2013htx}. We ascribe to $M_1$ a lower limit of 10~GeV, where DM would already be overabundant, and the upper limit of 350~GeV is chosen to match the $\mu$ upper bound, so that we can study light neutralino DM. In this case, DM can be either bino-like or higgsino-like, and mixed bino--higgsino LSP with mass $\sim$$\mathcal{O}(10~\mathrm{GeV})$ may also produce the right DM relic abundance.

We adopt the Markov Chain Monte Carlo (MCMC) method based on the Metropolis-Hastings algorithm to perform the scan of the MSSM parameter space with likelihood \mbox{$\propto$$\exp(-\chi^2_{\mathrm{tot}}/2)$}. The package \texttt{MicroMEGAS-5.2.30}~\cite{B_langer_2011} is used for DM relic density and DD cross section calculations, \texttt{SuperIso-4.0}~\cite{Mahmoudi:2008tp} is used for obtaining $B$-physics predictions, \texttt{HiggsTools}~\cite{Bahl:2022igd} (unification of \texttt{HiggsBounds-5}~\cite{Bechtle:2020pkv} and \texttt{HiggsSignals-2}~\cite{Bechtle:2020uwn}) is used to constrain SM-like and extra Higgs bosons at colliders. The total $\chi^2_{\mathrm{tot}}$ is defined as the sum of individual $\chi^2$ values of PLANCK relic density, $B$-physics, DM DD, and Higgs constraints from \texttt{HiggsSignals}:
\begin{equation}
    \chi^2_{\mathrm{tot}} =  \chi^2_{\Omega h^2} + \sum_i \chi^2_{B-\mathrm{physics}} + \chi^2_{\mathrm{DD}} + \chi^2_{\mathrm{HiggsSignals}}.
\end{equation} 

The $\chi^2$ of DM relic density is taken to be zero if the predicted value is below the central value of observed data of $0.1186 \pm 0.002$, and calculated with the formula
\begin{equation}
    \chi^2 = \frac{(\mu_t-\mu_0)^2}{\sigma_{\mathrm{theo}}^2+\sigma_{\mathrm{exp}}^2} \label{eq:chi2}
\end{equation}
otherwise, where $\mu_t$ is predicted from the theoretical value. $\mu_0$, $\sigma_{\mathrm{exp}}$ and $\sigma_{\mathrm{theo}}$ are experimental central value, we take the theoretical uncertainty of $0.1\mu_t$. Constraints from $B$-physics are Gaussian-distributed, the $\chi^2$ for each observable, take the form of Equation~\eqref{eq:chi2}. Experimental central value, experimental and theoretical uncertainty given in Table~\ref{tab:constraints}, and theoretical uncertainty of $0.1\mu_t$ is assumed if not provided in the table.

\begin{table}[htbp]
\centering
\caption{$B$-physics experimental data.}
\label{tab:constraints}
\begin{ruledtabular} %
\begin{tabular}{lcc}
\textbf{Observable} & \textbf{Value} & \textbf{Reference}\\
\hline %
${\rm BR}(B \rightarrow X_s \gamma)$ & $(3.27 \pm 0.14) \times 10^{-4}$ & ~\cite{GAMBITFlavourWorkgroup:2017dbx}  \\
${\rm BR}(B^0_s \rightarrow \mu^+ \mu^-)$ & $(3.34 \pm 0.27) \times 10^{-9}$ & ~\cite{ParticleDataGroup:2024cfk}\\
${\rm BR}(B^+ \rightarrow \tau^+ \nu_\tau)$ & $(1.09 \pm 0.24) \times 10^{-4}$ & ~\cite{ParticleDataGroup:2024cfk}\\
\end{tabular}
\end{ruledtabular} %
\end{table}

The estimation of $\chi^2$ for the DM-nucleus SI DD cross section $\chi^2_{\mathrm{DD}}$ is
\begin{equation}
    \chi^2_{\mathrm{DD}} = \left( \frac{\sigma^{\mathrm{SI}}_{\chi p}}{\sigma^{\mathrm{SI}, 90\%}_{\chi p}/1.64} \right)^2,
\end{equation}
where the effective theoretical cross section $\sigma^{\mathrm{SI}}_{\chi p}$ is strictly rescaled by the dark matter fractional abundance $f = \Omega_{\chi_1^0}h^2/0.118$ (assuming the local halo density fraction matches the cosmic abundance), and $\sigma^{\mathrm{SI}, 90\%}_{\chi p}$ represents the corresponding upper limit of the cross section for a given DM mass at 90\% confidence level from LZ~\cite{LZ:2024zvo}. By assuming null detection, we can take the central value as zero and the number 1.64 is the unit of 90\%~confidence level (C.L.)~\cite{Abdughani:2021oit}.

For the LHC constraints, we consider null results from SUSY searches with two or three leptons plus missing transverse momentum at the 13~TeV LHC with 36.1~fb$^{-1}$~\cite{ATLAS:2017mjy,ATLAS:2018ojr} and 139~fb$^{-1}$~\cite{ATLAS:2019lff,ATLAS:2019lng,ATLAS:2022zwa}. We also include projections for the 14~TeV HL-LHC with 3000~fb$^{-1}$~\cite{ATL-PHYS-PUB-2014-010}. Signal processes $pp \to \chi_1^+ \chi_1^-$ and $pp \to \chi_1^\pm \chi_{2,3}^0$ are generated using \texttt{MadGraph5\_aMC-v3.5.2}~\cite{Alwall:2014hca} with the default parton distribution function set~\cite{Buckley:2014ana} and rescaled with a $K$-factor of 1.5 to include next-to-leading corrections. Events are showered and hadronized with \texttt{PYTHIA-8.2}~\cite{Sjostrand:2014zea}, detector effects simulated with \texttt{DELPHES-3.5.0}~\cite{deFavereau:2013fsa}, and analyses recast with \texttt{CheckMATE-2.0.37}~\cite{checkmate2}. A sample is excluded at 95\% C.L. if the event ratio $r = \max(N_{S,i}/S^{95\%}_{\mathrm{obs},i})$ exceeds unity, where $N_{S,i}$ is the number of the events for the $i$-th signal region and $S^{95\%}$ is the corresponding observed 95\% C.L. upper limit.

We emphasize that the statistical framework utilized here corresponds to a profile likelihood approach, where the combined $\chi^2_{\mathrm{tot}}$ relates to the likelihood via $L \propto \exp(-\chi^2_{\mathrm{tot}}/2)$. The mathematically asymmetric treatment of the dark matter relic density—where the penalty is zero if the prediction is below the observed central value, and Gaussian if it exceeds it—is a strict physical requirement of this scenario. Because the neutralino is allowed to act as a subdominant relic within a multi-component dark matter universe, under-predicting the total dark matter abundance is phenomenologically valid, whereas over-predicting it is excluded. Thus, the relic density and direct detection limits are rigorously implemented as well-defined one-sided upper-bound likelihoods (half-Gaussians). Quantitatively, the survival of a parameter point is determined by the threshold $\chi_{\mathrm{tot}}^2 < 6$. This cutoff is adopted as a conservative benchmark tolerance to identify viable regions of the parameter space. It ensures that the selected points remain in reasonable overall agreement with the global combination of current cosmological, flavor, and collider constraints, without being strongly penalized or excluded by any individual measurement.

\section{Results} \label{sec:result}

\begin{figure}[htbp]
\includegraphics[width=0.48\linewidth]{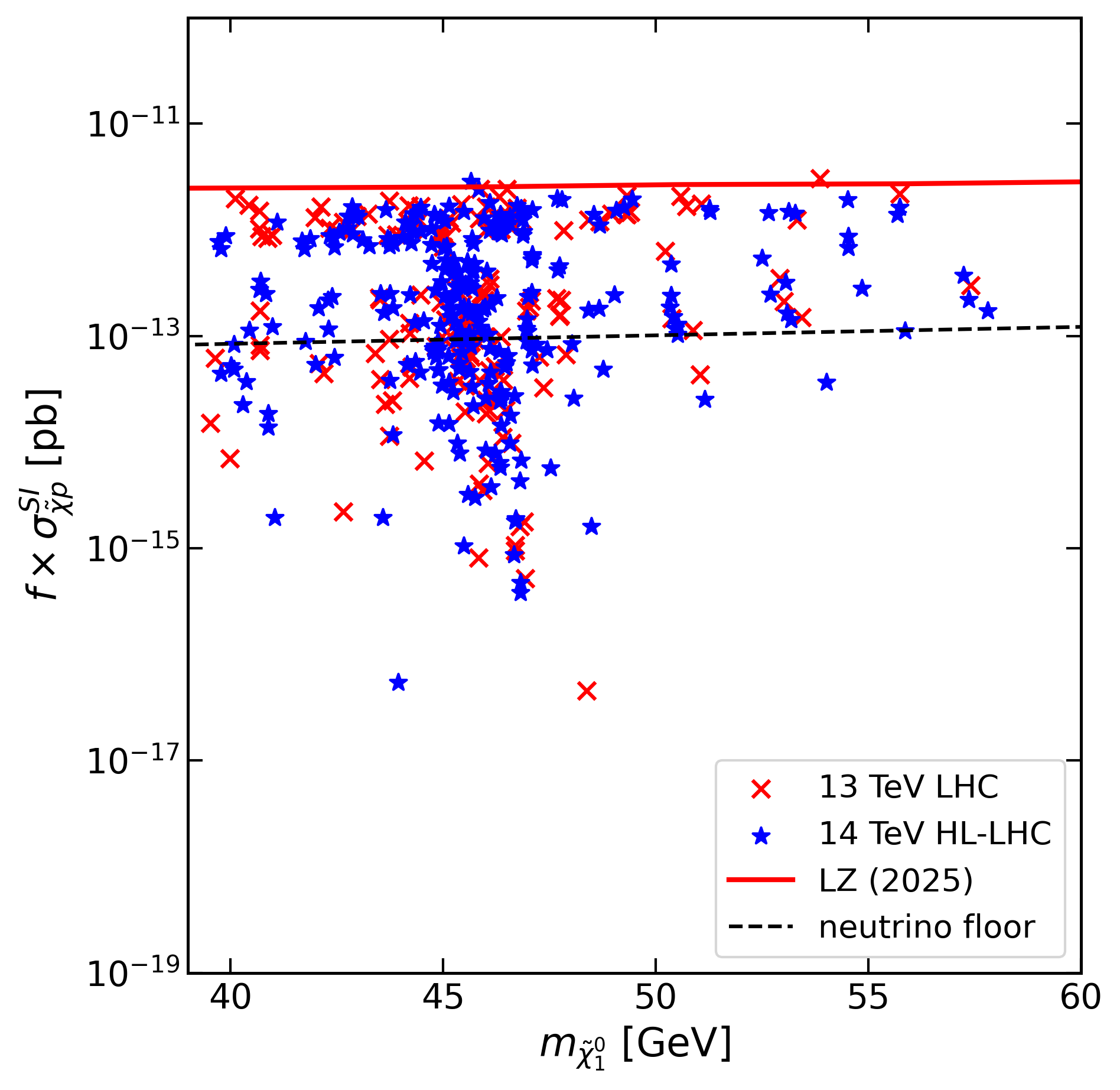}
\includegraphics[width=0.48\linewidth]{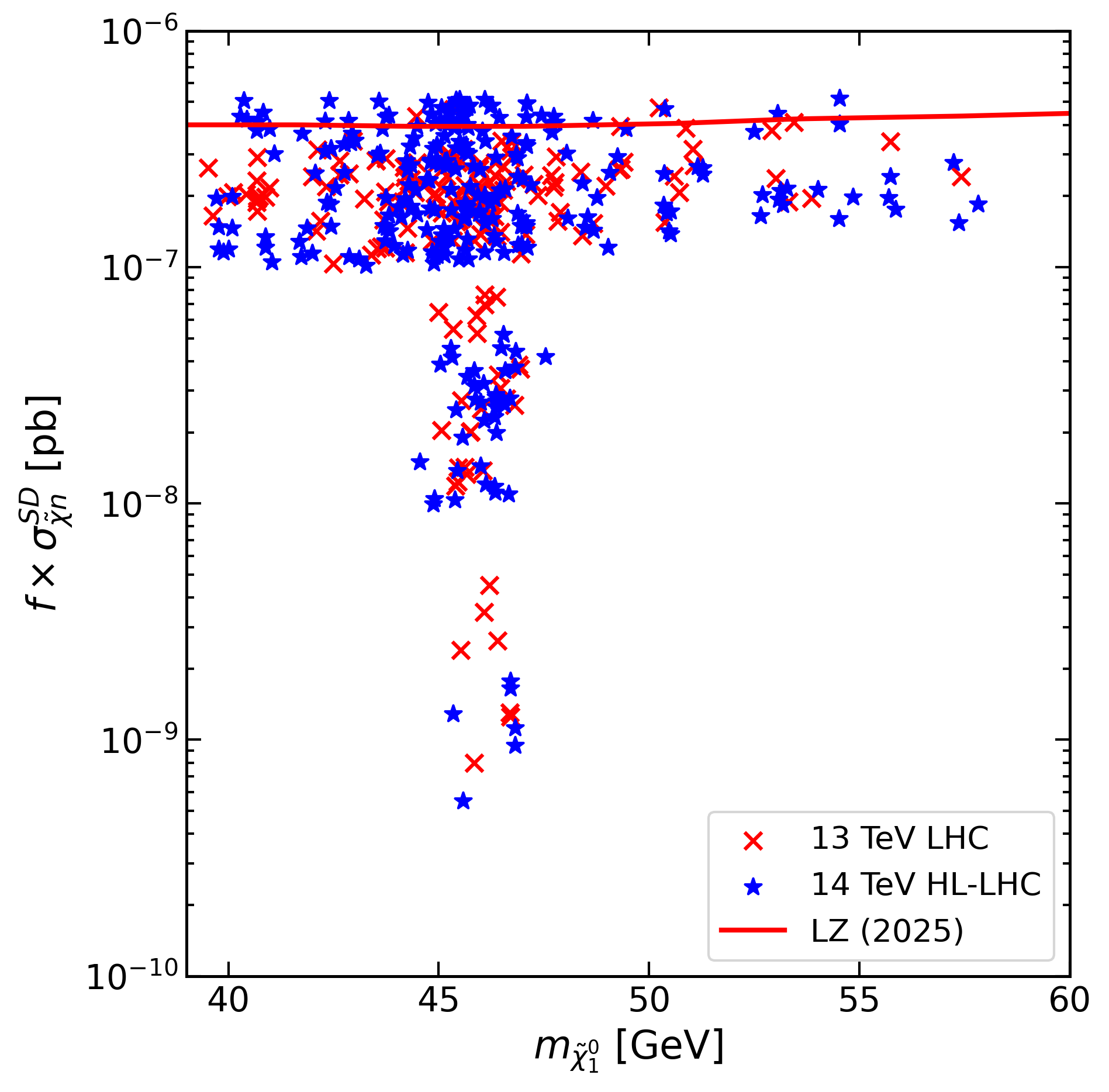}
\vspace{-3pt}
\caption{Spin-independent/dependent DM-nucleon cross sections. All samples satisfy the $\chi^2_{\mathrm{tot}} < 6$ constraint. Red crosses are excluded by current electroweakino searches at the 13~TeV LHC at \mbox{95\% C.L.}, while blue stars are projected to be searched at the 14~TeV HL-LHC with a luminosity of 3000~fb$^{-1}$. Red solid lines are 90\% C.L. upper limits from LZ (2025) experiments~\cite{LZ:2024zvo}. Black dashed line is the neutrino floor. $f$ represents the ratio of the predicted DM relic density to the value observed by Planck.}
\label{fig:DD}
\end{figure}

In Figure~\ref{fig:DD}, we show all samples satisfying the $\chi^2_{\mathrm{tot}} < 6$ limit in the SI and SD DM-nucleon cross section vs. DM mass space. $f$ indicates the ratio of the predicted DM relic density to the value observed by Planck, i.e., $f = \Omega_{\chi^0_1} h^2 / 0.118$. Red crosses are excluded by current electroweakino searches at the 13~TeV LHC at 95\% C.L., while blue stars are projected to be searched at the 14~TeV HL-LHC with the luminosity of 3000~fb$^{-1}$ at 95\% C.L. Red solid lines denote the 90\% C.L. upper limits from LZ (2025) experiments~\cite{LZ:2024zvo}, which gives the current strongest scattering cross section limit among the DD experiments at the DM mass of $\sim$10~GeV to 10~TeV range. Some points remain below the neutrino floor, beyond the reach of DD experiments. However, to evaluate the future collider reach, we explicitly recast the ATLAS HL-LHC projection analysis (ATL-PHYS-PUB-2014-010)~\cite{ATL-PHYS-PUB-2014-010} using \texttt{CheckMATE 2}. This simulation demonstrates that the next-generation HL-LHC with an integrated luminosity of 3000 fb$^{-1}$ will probe the surviving subdominant $Z$-resonance region within the natural SUSY parameter space defined by $\Delta_{\mathrm{EW}} < 30$ (corresponding to $|\mu| \lesssim 350$ GeV) and the chosen light-bino scan range.

\vspace{-3pt}

\begin{figure}[H]
\includegraphics[width=0.48\linewidth]{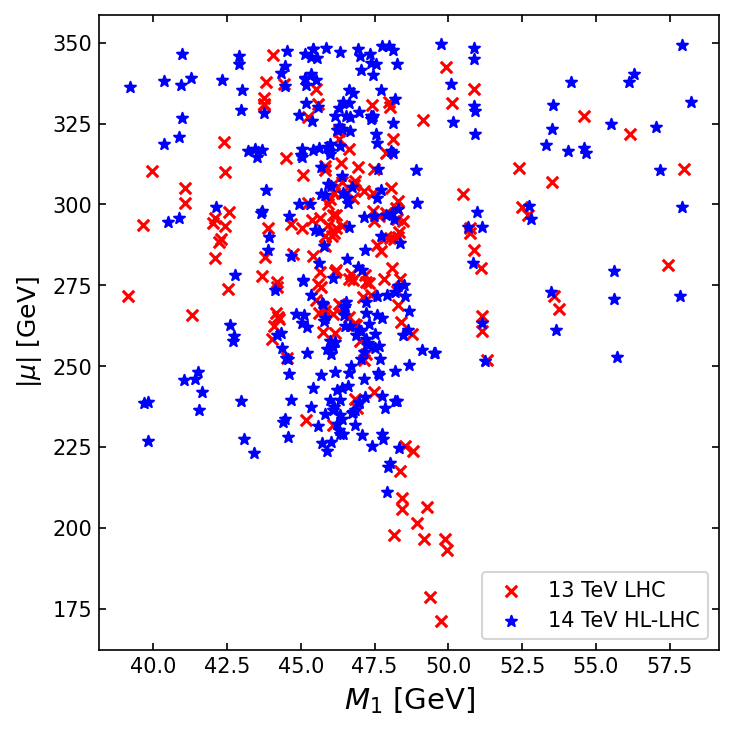}
\includegraphics[width=0.48\linewidth]{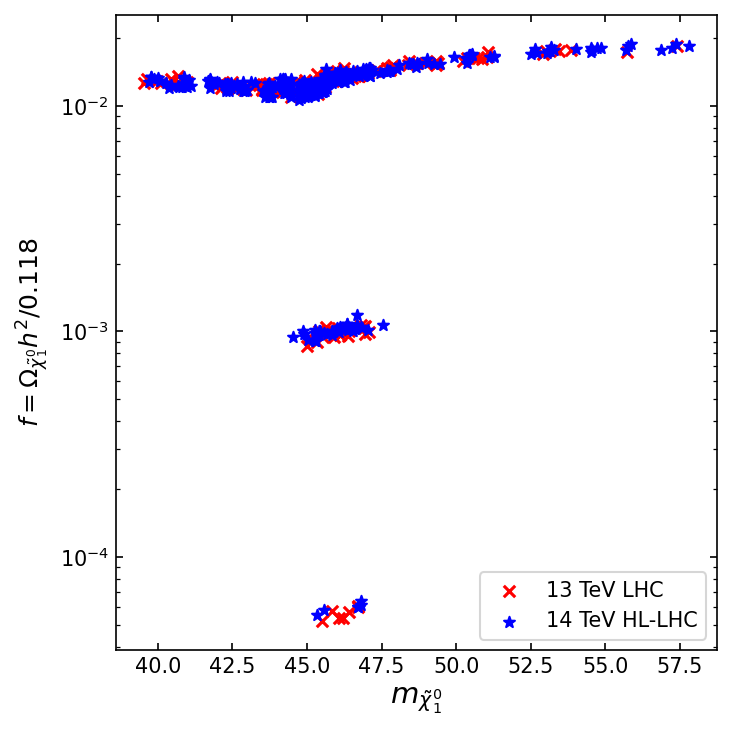}
\vspace{-3pt}
\caption{The same as Figure~\ref{fig:DD} but shown in the $\mu$ vs. $M_1$ plane (\textbf{left panel}) and the $f$ vs. $m_{\tilde{\chi}_1^0}$ plane (\textbf{right panel}). Here, $f$ denotes the ratio of the predicted dark matter relic density to the value observed b\mbox{y Pl}anck.}
\label{fig:mu}
\end{figure}

The left panel of Figure~\ref{fig:mu} shows the sample on the MSSM soft parameters $\mu$ vs. $M_1$ space. Requiring $\chi^2_{\mathrm{tot}} < 6$ constrains $M_1$ to lie near the $Z$-resonance, while the lower bound on $|\mu|$ is $\sim$170~GeV. Current 13~TeV LHC data raise this limit to $\sim$200~GeV. Contributions to the $\chi^2_{\mathrm{tot}}$ mainly come from strong LZ (2025) DD limit, and it pushes the coupling $g_{h\chi_1^0 \chi_1^0}$ to the small enough value so that Higgs resonance is excluded. 

It is highly instructive to compare these findings with recent comprehensive scans of the broader phenomenological MSSM (pMSSM) parameter space~\cite{Barman:2024}. In their broad analysis, which included varying light sfermion masses, the authors similarly found that current collider and direct detection data heavily restrict the light neutralino to the $Z$-resonance region as a subdominant relic. The convergence of our results with theirs is phenomenologically significant. While their study employed a bottom-up scanning approach, our framework is driven by strict top-down naturalness requirements ($\Delta_{\mathrm{EW}} < 30$ and decoupled 3~TeV sfermions). Furthermore, by extending our scan up to $M_1 = 350$~GeV, we explicitly demonstrate that heavier bino masses fail to survive the stringent 2025 LZ limits. Thus, our results corroborate the pMSSM findings while proving that, within the strict confines of Natural SUSY, this heavily squeezed, subdominant $Z$-resonance is the only remaining viable regime for light neutralino dark matter. We further note a significant cosmological consequence of these constraints: the complete exclusion of the bino--higgsino coannihilation scenario within this framework. In the surviving $Z$-resonance parameter space, the strict requirement of $M_1$$\sim$45~GeV and $|\mu| \gtrsim 170$~GeV enforces a large mass splitting between the bino-like LSP and the higgsino-like NLSPs ($\Delta m \gtrsim 125$~GeV). With such a substantial mass gap, the higgsino states have effectively decayed away before the bino freeze-out, meaning coannihilation processes play no role in determining the relic density. The compressed spectrum regions ($M_1$$\sim$$\mu$) where bino--higgsino coannihilation would typically dominate are definitively ruled out by the 2025 LZ spin-independent limits due to the large neutralino mixing in those regimes. Consequently, the relic density in the surviving parameter space is driven entirely by resonant annihilation through the $Z$ boson.

Previously, theoretical blind spots~\cite{Abdughani:2017dqs}—where a negative relative sign ($M_1/\mu < 0$) induces a tuned cancellation in the $g_{h\chi_1^0\chi_1^0}$ coupling—allowed the $h$-resonance region to evade direct detection. However, our updated analysis reveals that the 2025 LZ limits are now so overwhelmingly stringent that even these tuned blind spot configurations are completely excluded, entirely ruling out the $h$-resonance region. Consequently, the survival of the remaining parameter space, strictly localized to the $Z$-resonance, does not rely on tuned coupling cancellations. Both the small spin-independent (SI) and spin-dependent (SD) effective cross sections in this surviving region are solely the result of the deeply suppressed fractional abundance ($f$). Because resonant annihilation at the $Z$-pole is highly efficient, the dark matter fraction drops significantly, allowing the effective scattering rate ($f \times \sigma^{\mathrm{SI}, \mathrm{SD}}$) to safely fall below current LZ sensitivities.

On the right panel of Figure~\ref{fig:mu}, an upper limit of $\sim$0.02 and a lower limit of $\sim$$5 \times 10^{-5}$ have been obtained for $f$, i.e., the bino-like DM in this scenario can constitute at most 2\% of the total DM relic. While recent comprehensive pMSSM scans~\cite{Barman:2024} have similarly identified the survival of this $Z$-resonance region in the mass parameter space, they do not explicitly quantify the resulting relic density. Our analysis complements those broader exclusions by explicitly demonstrating the profound suppression of the dark matter fraction in this regime. We further note that the surviving samples cluster into distinct horizontal bands. The upper band represents the parameter boundary enforced by the stringent LZ limits, while the lowest band corresponds to points sitting exactly on the peak of the $Z$-resonance ($m_{\tilde{\chi}_1^0} \approx m_Z/2$), where the annihilation cross-section is maximized. The concentration into these discrete bands, rather than a continuous distribution, is an artifact of the Markov Chain Monte Carlo (MCMC) sampling algorithm; the MCMC walkers preferentially cluster at these distinct local likelihood maxima (the exact resonance peak and specific viable mixing plateaus) rather than uniformly populating the extremely steep and narrow slopes of the $Z$-resonance.

In addition, recent analyses of the Muon $g-2$ experiment at Fermilab report the value of $a_\mu^{\mathrm{exp}} = 1165920715(145) \times 10^{-12}$~\cite{Muong-2:2025xyk}, while theoretical SM prediction is \mbox{$a_\mu^{\mathrm{SM}} = 116592033(62) \times 10^{-11}$~\cite{Aliberti:2025beg}}. This suggests that no significant discrepancy remains between experiment and theory. Regarding the muon anomalous magnetic moment, we find that the electroweakino contribution in our surviving parameter space is tightly constrained to $\mathcal{O}(10^{-11})$. In accordance with previous results, widely discussed in the literature, in similar low-mass neutralino regimes ~\cite{Barman:2024}.

Given these profound constraints on the relic density, it is crucial to establish the cosmological setup necessitated by this experimental landscape. Because our analysis demonstrates that the light bino--higgsino neutralino contributes at most $\sim$2\% observed Planck relic density, this framework can no longer be interpreted as a standalone dark matter model but must instead be viewed strictly as a subdominant relic within a multi-component dark matter scenario. In such a setup, the remaining bulk of the dark matter abundance must originate from another source. Within the supersymmetric framework, a highly compelling candidate for this dominant component is the axion, which naturally emerges from the Peccei–Quinn mechanism introduced to solve the strong CP prob\mbox{lem~\cite{pq1,Pq,axion,axion2}}. While our results demonstrate that the neutralino can only make up a small fraction of the cosmic abundance, it uniquely retains electroweak interactions. Therefore, this subdominant component still entirely dictates the observable phenomenology at colliders and WIMP-nucleon direct detection experiments, whereas probing the dominant axion component relies on orthogonal experimental strategies, such as haloscopes~\cite{halo,halo1}.

\section{Conclusion}
\label{sec:conclusion}

We have examined the light bino--higgsino dark matter scenario in the framework of natural SUSY, motivated by both electroweak naturalness and the latest experimental results. Requiring $\Delta_{\mathrm{EW}} < 30$ restricts the higgsino mass parameter to $|\mu| \in [100,350]$~GeV, while we extended the bino mass to $M_1 \in [10,350]$~GeV. After imposing constraints from Higgs data, rare $B$ decays, LEP limits, direct detection searches, and the observed relic density, we find that only a small fraction of the parameter space remains viable. In particular, the neutralino relic density is always well below the Planck value, contributing at most $\sim$2\% of the total dark matter abundance.

Our analysis shows that current 13 TeV LHC electroweakino searches already exclude part of the parameter space, while the future HL-LHC with 3000 fb$^{-1}$ luminosity will be able to probe the remaining region under the imposed $\Delta_{EW}<30$ naturalness condition and the chosen scan range. Direct detection experiments such as LZ have already reached strong sensitivity, and future improvements may probe most of the surviving points, except for those lying below the neutrino floor. Moreover, in this scenario, the electroweakino contribution to the muon $g-2$ is of order $10^{-11}$, consistent with the latest Fermilab measurement and its reduced discrepancy with the Standard Model prediction.

\section*{Acknowledgements}

This research was funded by the National Natural Science Foundation of the People's Republic of China grant number 12303002 and Tianchi talent project of Xinjiang Uygur Autonomous Region of China.

\vspace{8pt}

\bibliography{references}
\end{document}